\documentclass[twocolumn,showpacs,amsmath,amssymb,pra,floatfix]{revtex4}

\usepackage{graphicx}
\usepackage{dcolumn}
\usepackage{bm}
\usepackage{bbm}
\usepackage{multirow,amssymb,amsbsy,amsmath}
\usepackage{stmaryrd}
\usepackage{color}

\newcommand{\ddim}{\udelta\kern0.1em}

\newcommand{\beikonst}[2]{\left( #1 \right)_{\kern-0.2em #2}}

\newcommand*{\ket}[1]{\mathopen{|}#1\mathclose{\rangle}}

\begin{document}


%
%

\title{String order in dipole-blockaded quantum liquids}

\author{Hendrik Weimer}%
\email{hweimer@itp.uni-hannover.de}
\affiliation{Institut f\"ur Theoretische Physik, Leibniz Universit\"at Hannover, Appelstr. 2, 30167 Hannover, Germany}

\date{\today}%

\begin{abstract}

  We study the quantum melting of quasi-one-dimensional lattice
  models in which the dominant energy scale is given by a repulsive
  dipolar interaction. By constructing an effective low-energy theory,
  we show that the melting of crystalline phases can occur into two
  distinct liquid phases, having the same algebraic decay of
  density-density correlations, but showing a different non-local
  correlation function expressing string order. We present possible
  experimental realizations using ultracold atoms and molecules,
  introducing an implementation based on resonantly driven Rydberg
  atoms that offers additional benefits compared to a weak admixture
  of the Rydberg state.

\end{abstract}


\maketitle

\section{Introduction}

The constrained scattering in one-dimensional (1D) quantum systems
allows for their effective description in terms universal low-energy
theories even when the microscopic model is not exactly solvable
\cite{Giamarchi2004}. The most prominent example is the Luttinger
liquid, in which all correlation functions decay algebraically
according to a single parameter \cite{Haldane1981}. However, the
relation between the actual particles of interest and the low-energy
quasiparticles is not always trivial. In this article, we show that for
quantum liquids with dominant long-range interactions, the
transformation between the two can be highly nonlocal, giving rise to
a quantum phase transition between Luttinger liquids differing by
string order. Realizing and understanding such nonlocal or topological
order is of immense interest as it serves as a key stone to develop a
more general theory of phase transitions beyond the Landau symmetry
breaking paradigm \cite{Wen2013}.

The observation of such exotic phase transitions is often tied to the
presence of strong tunable interactions, hence dipolar interactions
found within polar molecules \cite{Carr2009,Baranov2012} or Rydberg
atoms \cite{Saffman2010,Low2012} serve as ideal candidates and also
allow for the combination with well-established tools for studying 1D
physics within ultracold quantum gases
\cite{Paredes2004,Kinoshita2004,Syassen2008,Haller2010,Simon2011,Endres2011}. These
recent developments have led to to a wide range of theoretical studies
investigating the ground state properties of dipoles in 1D
\cite{DallaTorre2006,Kollath2008,Burnell2009,Schachenmayer2010,Hauke2010,Pikovski2010,Dalmonte2010,Weimer2010a,Sela2011,Lauer2012,Ruhman2012,Bauer2012,Knap2012,Manmana2013,Mattioli2013,Gammelmark2013},
giving rise to a plethora of novel many-body phenomena. Of particular
interest is the regime of strong repulsive interactions, in which the
dipole blockade excludes configurations having two particles in close
proximity and leads to strong frustration effects. In the absence of
quantum fluctuations, the ground state of a dipole-blockaded lattice
gas is characterized by a devil's staircase of gapped crystalline
phases commensurate with the underlying lattice
\cite{Bak1982}. Generically, the quantum fluctuations induced by
movement of the particles result in commensurate-incommensurate
transitions to a Luttinger liquid
\cite{Burnell2009,Sela2011,Lauer2012}. Additional phases can occur
pertaining to extended interaction potentials \cite{Mattioli2013} or
quasi-1D geometries \cite{Bauer2012,Gammelmark2013}.
\begin{figure}[t]

\includegraphics{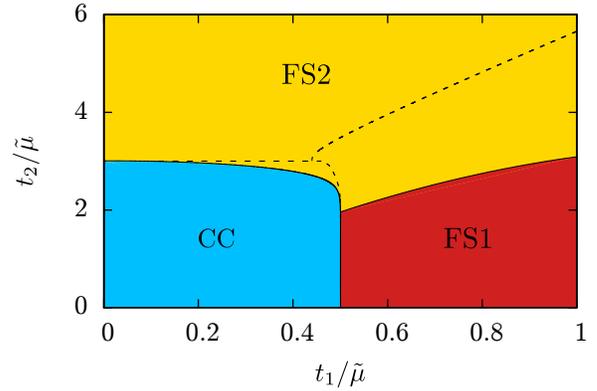}

\caption{Ground state phase diagram ($U=4\tilde{\mu}$). Melting of the
  commensurate crystal (CC) induced by nearest-neighbor hopping $t_1$
  and next-nearest neighbor hopping $t_2$ result in two distinct
  floating solid phases (FS1 and FS2) differing by a non-local
  operator characterizing string order. The dashed lines correspond to
  predictions from a mean-field treatment of the low-energy
  theory.}

\label{fig:phase}

\end{figure}

In this article, we build on these earlier developments and study
dipole-blockaded quantum gases on a triangular ladder. We establish
the ground state phase diagram by analyzing an effective low-energy
theory describing the dynamics of dislocation defects of the
commensurate crystals. Crucially, the melting of the commensurate
crystals can be induced by motion either along the direction of the
ladder or along its rungs. This leads to the appearance of two
distinct floating solid phases, see Fig.~\ref{fig:phase}, both of
which can be described in terms of a Luttinger liquid. Remarkably, we
find that the two floating solids cannot be distinguished by merely
looking at correlation functions of local operators; instead one has
to consider a highly non-local observable describing string
order. Finally, we comment on possible experimental realizations using
ultracold polar molecules or Rydberg atoms, including a novel approach
for the latter using laser-induced hopping of Rydberg excitations in
an electric field gradient, which can be also used to implement a
large class of microscopic models with an unprecedented level of
control over hopping and interaction parameters.

\section{Hamiltonian description}

\begin{figure}[t]
  \includegraphics[width=\linewidth]{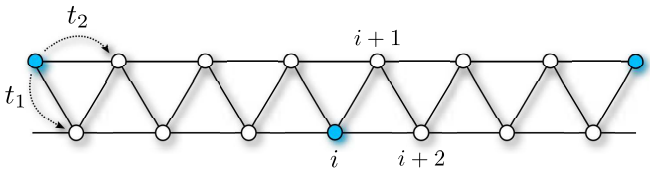}
  \caption{Setup of the system. Dipolar particles are confined to a
    triangular ladder structure, with hopping occuring along the
    direction of the ladder ($t_2$) or along its rungs ($t_1$). Filled
    dots indicate the particle positions corresponding to the $q=7$
    commensurate crystal.}
\label{fig:setup}
\end{figure}

We start our analysis based on the microscopic Hamiltonian in terms of
an extended Hubbard model with long-range dipolar interactions, with
the setup of the system depicted in Fig.~\ref{fig:setup}. In the
following, we treat the triangular ladder as a single chain having
nearest and next-nearest neighbor hoppings. We point out that the
dipole blockade renders the distinction between bosons and fermions
irrelevant as the exchange of two particles occurs at very high energy
scales, which are unimportant for the low-energy properties of the
system. The Hamiltonian is given by
\begin{eqnarray}
  H &=& -t_1 \sum\limits_i \left(c_i c^\dagger_{i+1} + \mathrm{H.c.}\right) -t_2 \sum\limits_i \left(c_i c^\dagger_{i+2} + \mathrm{H.c.}\right)\nonumber\\ &+& \sum\limits_{i<j} V_{|i-j|} n_in_j  - \mu \sum\limits_i n_i.
\label{eq:H}
\end{eqnarray}
Here, $t_1$ and $t_2$ are the strength of the nearest and next-nearest
neighbor hopping, respectively, $V_{|i-j|}$ accounts for the repulsive
dipolar interaction between sites $i$ and $j$ according to the
particle number $n_i=c_i^\dagger c_i$, and $\mu$ denotes the chemical
potential. In the classical limit with $t_1=t_2=0$, the ground state
again follows a complete devil's staircase structure of commensurate
crystals as the interaction potential is a convex function
\cite{Bak1982}. The most stable commensurate crystals occur at
rational fillings $1/q$ with $q$ being odd, i.e., the particles are
located on the two legs of the ladder in an alternating fashion, see
Fig.\ref{fig:setup}.  In the following, we will restrict our analysis
to densities close to these values. Here, we are interested in the
dipole-blockaded regime with $q\gg 1$, which allows us to approximate
many quantities of interest by performing expansions in $1/q$
\cite{Weimer2010a}. For example, the center of the commensurate
crystals with filling $1/q$ occurs at a chemical potential of $\mu_0
\approx 32\zeta(3)V_1/q^3$, and the variation in chemical potential
over which the phase is stable is given by $\mu_w \approx
168\zeta(5)V_1/q^4$.

\section{Effective low-energy theory} 

We now study the effects of quantum fluctuations induced by $t_1$ and
$t_2$ within perturbation theory, i.e., $t_1,t_2\ll \mu$
\cite{Burnell2009,Weimer2010a}. The low-energy excitations correspond
to dislocation defects of the commensurate crystal, given by the
relation $d_j = r_{j+1}-r_j-q$, which measures the deviation of the
spacing between the particles $j$ and $j+1$ from the perfectly
commensurate case. Note that these defects are nonlocal
quasiparticles, as their position in terms of the original particles
depends on the number of defects located at previous
sites. Consequently, changing the notation from real particles to
defects is a highly nonlocal transformation. To denote this crucial
distinction between lattice sites and defects, we will use the index
$i$ when referring to the former and $j$ for the latter. Depending on
the sign of $d_j$, defects occur as hole-like or particle-like, i.e.,
they decrease or increase the total density, respectively. However,
sufficiently far away from the particle-hole symmetric point given by
$\mu = \mu_0$, only one of these defects is relevant
\cite{Weimer2010a}. Furthermore, the number of defects is a conserved
quantity. As the energy cost rapidly increases for $|d_j| > 1$, and
the hopping of defects does not exhibit bosonic enhancement, we
restrict the Hilbert space to the defect numbers $d_j = 0,1,2$. Then,
the effective low-energy Hamiltonian to first order in $t_1$, $t_2$,
can be expressed using spin-$1$ variables as
\begin{eqnarray}
H &=& -t_1 \sum_j \left(S_j^+S_{j+1}^- + \mathrm{H.c}\right)\nonumber\\ &-& t_2 \sum_j \left(S_j^+S_j^+S_{j+1}^-S_{j+1}^- + \mathrm{H.c.}\right) \nonumber\\ &+& \tilde{\mu} \sum_j \left(1+S_j^z\right)+ U \sum_j S_j^+S_j^+S_j^-S_j^-,
\label{eq:Hd}
\end{eqnarray}
i.e., the next-nearest neighbor hopping $t_2$ turns into a correlated
hopping of the defects. Most importantly, the strong dipolar
interaction has been absorbed into the definition of the defects;
hence, the resulting low-energy Hamiltonian is purely local and can be
further analyzed using standard techniques. In addition to higher
order processes in the perturbation series, we also neglect the weak
interaction between the defects. The energy cost associated with each
defect is given by $\tilde{\mu}=(\mu_0-\mu+\mu_w/2)/q$, and the
repulsion of the defects can be calculated as $U=\mu_w/q$. Note that
this model is equivalent to a Bose-Hubbard model with correlated
hopping and a three-body constraint \cite{Mazza2010}.

\begin{figure}[t]

\includegraphics{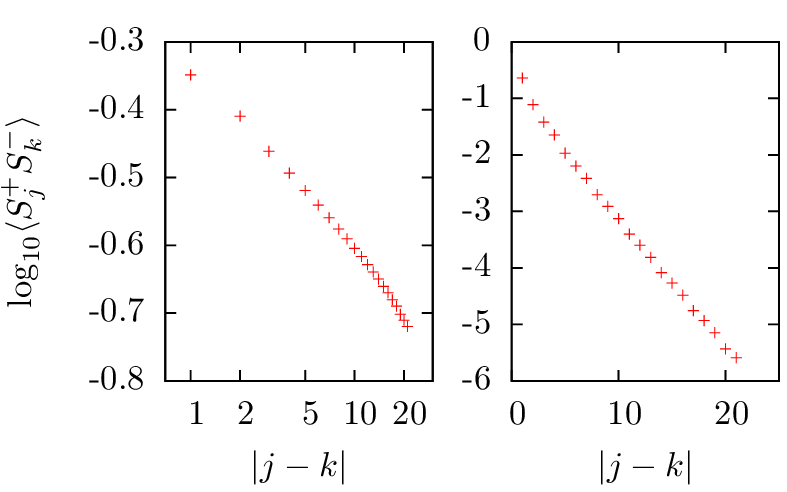}

\caption{Decay of spin correlations in the low-energy theory for a
  system of 30 spins ($U=4\tilde{\mu}$, $t_1=0.6\tilde{\mu}$). Upon
  increasing the pair hopping $t_2$, the system undergoes a phase
  transition from an atomic defect liquid with an algebraic decay
  ($t_2 = 1.5\tilde{\mu}$, left) to a pair defect liquid showing an
  exponential decay ($t_2 = 3.0\tilde{\mu}$, right).}

\label{fig:decay}

\end{figure}

If one of the hopping term vanishes, the phase boundaries can be
determined exactly by mapping the problem onto free fermions
\cite{Sachdev1999}. For $t_2=0$, the on-site repulsion $U$ is
irrelevant at the phase transition, which occurs at $t_1 =
\tilde{\mu}/2$ between the $n=0$ Mott insulator and a liquid phase
with finite defect density. Likewise, there is a second phase
transition for $t_1=0$ occuring at $t_2 =
\tilde{\mu}+U/2$. Remarkably, this second liquid has defects always
appear in pairs as the single-defect sector is still protected by a
gap of $\tilde{\mu}$. In the following, we refer to the latter phase
as a ``pair defect liquid'', while calling the former a ``single
defect liquid''. Based on the low-energy Hamiltonian,
eq.~(\ref{eq:Hd}), we map out the complete phase diagram using
mean-field theory and an exact density-matrix renormalization group
(DMRG) method based on a matrix-product state approach
\cite{Wall2014,Wall2012}, see Fig.~\ref{fig:phase}. The transition
line between the two liquid phases corresponds to the decay of the
spin correlation function $\langle S^+_jS^-_k\rangle$ changing from
algebraic to exponential behavior, as shown in
Fig.~\ref{fig:decay}. Here, we determine the transition line from the
comparison of an algebraic and an exponential fit to the correlation
function for a system of 30 spins. As noted previously
\cite{Mazza2010}, mean-field theory produces good qualitative
agreement with the DMRG results, and furthermore yields the correct
values for the transition in the exactly solvable cases.

In order to understand the transition between the two liquid phases in
more detail, it is instructive to represent the effective spin-$1$
model by two spin-$1/2$ degrees of freedom, which then can be
bosonized \cite{Timonen1985}. Then, sufficiently far away from the
transition, we know from the free fermion solution that the system is
well described in terms of a single component Luttinger liquid, i.e.,
the second bosonic field is massive, according to the effective
Hamiltonian
\begin{equation}
  H = \frac{v_j}{2\pi}\int \mathrm{d}x [K_j(\pi\Pi_j(x))^2 +\frac{1}{K_j}(\nabla\phi_j(x))^2],
\end{equation}
where $\Pi_j$ and $\nabla\phi_j$ are bosonic fields corresponding to
phase and density fluctuations. If the fields with $j=1$ are gapless,
then the system is in the single defect liquid phase, while gapless
$j=2$ fields correspond to the pair defect liquid. In the limit of low
defect densities, we find $K_j=K=1$ for the Luttinger parameter, while
the speed of sound is given by $v_1 = qa\sqrt{\tilde{\mu}t_1/2}$ and
$v_2 = qa\sqrt{(2\tilde{\mu}+U)t_2/2}$, respectively. The transition
between the single and the double defect liquid is of the Ising
universality class \cite{Romans2004,Manmana2011,Ejima2011}. From the
finite-size scaling behavior of the underlying Ising transition
\cite{Burkhardt1985}, we can estimate the error in determining the
phase boundary between the two liquid phases in our DMRG calculation
to behave as $\sim 1/\tilde{L}^2$, with $\tilde{L}$ being the number
of bulk spins considered in the fitting procedure. Here, we have used
a value of $\tilde{L}=18$, corresponding to an error from the
finiteness of the system of about one percent.

\section{String order} 

Within the validity of our perturbative approach, the phase boundaries
of the defect model (\ref{eq:Hd}) corresponds to the phase boundaries
of the microscopic Hamiltonian (\ref{eq:H}). However, we are rather
interested in describing the appearing quantum phases in terms of
observables involving the microscopic degrees of freedom, i.e.,
correlations between individual particles rather than correlations
between the defects. In the following, we apply Luttinger liquid
theory to classify the ground state phases in terms of the microscopic
particles.

When mapping from the defect description to the real particles, we
first note that the $n=0$ Mott insulator for the defects corresponds
to the commensurate crystal at filling $1/q$, in which the
density-density correlation $\langle n_{iq}n_0\rangle$ exhibits true long-range
order. In the two liquid phases, we find the density-density
correlations of the microscopic particles to asymptotically decay as $
\langle n_xn_0 \rangle \sim~x^{-2K/(n_d+q)^2}$, where $n_d$ is the
density of the defects \cite{Weimer2010a}. Consequently, while the
existence of algebraically decaying correlations signals the melting
of the commensurate crystal phase, it is not possible to distinguish
the two defect liquids. Thus, explaining the phase diagram in terms of
the microscopic particles requires the probing of nonlocal
correlations. However, we already know that the single defect
correlation $\langle(1-S_z^2)^{(0)} (1-S_z^2)^{(j)} \rangle$ exhibits
an algebraic decay in the single defect liquid, and an exponential
decay in the pair defect liquid. Remarkably, here we find that this
behavior can be captured in terms of the microscopic variables
by introducing an observable measuring string order,
\begin{equation}
  O_\mathrm{string}(x) = \left\langle \exp\left(i2\pi/q\sum_{k=0}^x n_kn_{k+q-1}\right)\right\rangle.
\end{equation}
Here, we have focused on the case of particle-like defects, an
analogous expression for hole-like defects follows by replacing
$n_{k+q-1}$ by $n_{k+q+1}$. Most importantly, the term inside the
exponential is proportional to the number of single defects $N_x$
occuring over a distance $x$. Then, the value of
$O_\mathrm{string}(x)$ simply follows from the characteristic function
of the probability distribution of $N_x$. In the pair defect liquid
phase, the single defects are uncorrelated, meaning $N_x$ satisfies a
Poisson distribution with a mean growing linearly with
$x$. Consequently, $O_\mathrm{string}(x)$ decays exponentially with
distance in the pair liquid phase. In the single defect liquid,
however, $N_x$ is given by a discrete Gaussian distribution whose mean
also grows linearly with $x$, but having a variance $\sigma^2 =
K\log(x/b)/\pi^2$, where $b$ is a short distance cutoff
\cite{Weimer2010a}. From its characteristic function, we identify the
leading term in the long distance limit decaying according to an
algebraic function, $O_\mathrm{string}(x) \sim x^{-2K/q^2}$.

As the slowest decaying correlation function is still given by the
microscopic density-density correlations, both phases form a
``floating solid'' on top of the underlying lattice. We denote them by
FS1 and FS2, respectively, with the former corresponding to the single
defect liquid and thus exhibiting an algebraic decay of the string
correlations. Note that in contrast to the phases exhibiting string
order known as Haldane insulators
\cite{denNijs1989,Kennedy1992,DallaTorre2006}, both floating solid
phases are gapless.  The full phase diagram is shown in
Fig.~\ref{fig:phase}.

\section{Experimental realization}

Let us now turn to possible experimental implementations of the
extended Hubbard model introduced in Eq.~(\ref{eq:H}). In any of the
setups discussed in the following, the triangular lattice structure is
created using standard optical lattice beams
\cite{Bloch2008}. Additionally, string order can be measured by direct
imaging of atoms or molecules in the lattice
\cite{Simon2011,Endres2011}.

\subsection{Ultracold polar molecules}

As a first possible implementation, we consider a setup based on
ultracold polar molecules \cite{Carr2009,Baranov2012}. Here, the
molecules are prepared in the rovibrational ground state and loaded
into the triangular lattice. The hopping matrix elements $t_1$ and
$t_2$ follow from the tunneling of the molecules in the lattice
potential. The repulsive dipole-dipole interaction $V_{ij}$ can be
realized either by applying a strong electric field \cite{Buchler2007}
or by microwave dressing of the rotational excitations
\cite{Lemeshko2012,Yan2013}. For LiCs molecules having an electric
dipole moment of $d = 5.5\,\mathrm{D}$, the characteristic energy
scale $\tilde{\mu}$ close to the $q=7$ commensurate crystal on a $a =
532\,\mathrm{nm}$ lattice is given by $\tilde{\mu} \approx 2\pi\hbar \times
100\,\mathrm{Hz}$, which is compatible with experimental timescales
within these systems.

\subsection{Rydberg atoms}

\begin{figure}[b]
  \includegraphics[width=0.9\linewidth]{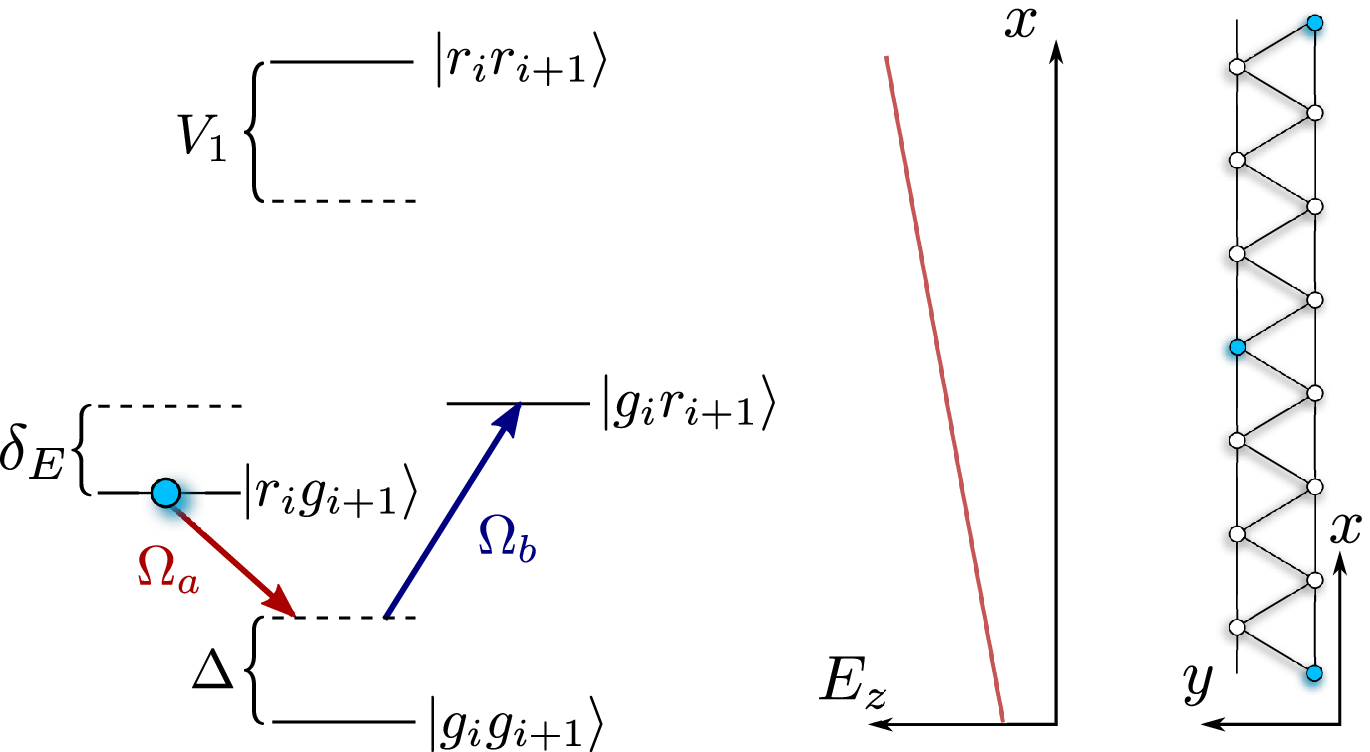}
  \caption{Energy levels of two adjacent atoms for laser-induced
    hopping of Rydberg excitations in an electric field gradient. The
    detuning between two Rydberg excitation lasers compensates the
    differential Stark shift $\delta_E = d(E_z^{(i+1)}-E_z^{(i)})$ created by
    the field gradient, while the $\ket{r_ir_{i+1}}$ state becomes far
    detuned through the dipolar interaction $V_1$.}
\label{fig:Egrad}
\end{figure}
Alternatively, our model can also be realized using ultracold Rydberg
atoms \cite{Saffman2010,Low2012}. A straightforward implementation
would consist of a weak coupling $\Omega$ to a Rydberg state detuned
by $\Delta_r$ \cite{Henkel2010,Pupillo2010,Honer2010}, where the
strong repulsive interactions between Rydberg states create an
interaction potential asymptotically decaying as $1/x^3$ for Rydberg
states within the Stark fan. However, the experimental parameters for
such a Rydberg dressing are quite challenging: in particular, the
dipolar interaction is suppressed by a factor $\sim
(\Omega/\Delta_r)^4$, while the radiative decay limiting the lifetime
of the system only decreases as $(\Omega/\Delta_r)^2$. Therefore, we
present here a different route benefitting from resonant excitations
to the Rydberg state. Initially, the atoms are loaded into a deep
optical lattice, forming a Mott insulator state with one atom per
lattice site. Then, the extended Hubbard model defined in
Eq.~(\ref{eq:H}) is realized by treating atoms in their electronic
ground state $\ket{g}$ as empty sites, and atoms in a Rydberg state
$\ket{r}$ as particles. Here, a finite density of Rydberg excitations
is created by adiabatically tuning the excitation lasers
\cite{Pohl2010,vanBijnen2011,Weimer2012}, which will control the value
of the chemical potential $\mu$. Finally, an electric field gradient
is introduced, such that the difference in the Stark shift between
different sites is exactly canceled by the detuning between two
excitation lasers, see Fig.~\ref{fig:Egrad}, resulting in a hopping of
the Rydberg excitations. Note that this process crucially relies on
the dipole blockade between neighboring sites; for non-interacting
particles the two paths via $\ket{g_ig_{i+1}}$ and $\ket{r_ir_{i+1}}$
interfere destructively. By introducing an additional laser, it is
possible to satisfy this resonance condition for both nearest-neighbor
and next-nearest-neighbor distances. The coupling constants $t_1$ and
$t_2$ derived from the induced hoppings of the Rydberg excitations
$\sim \Omega_a\Omega_b/2\Delta$ can be controlled independently by the
intensities of the excitation lasers. Here, we find that for a Rydberg
state with a principal quantum number of $n=43$ in an $a =
1\,\mathrm{\mu m}$ lattice, the liquid phases close to the $q=7$
commensurate crystal form around a characteristic energy scale of
$\tilde{\mu} \approx 2\pi\hbar \times 400\,\mathrm{kHz}$, which is
several orders of magnitude larger than the decay rate of the Rydberg
state. We would like to stress that this implementation procedure
based on electric field gradients is quite general and can readily be
extended to a large class of extended Hubbard models with tunable
long-range hoppings and interactions.

\section{Conclusions}

In summary, we have shown that dipole-blockaded quantum gases on
triangular ladders support two distinct liquid phases, differing by
string order. While the identical behavior of local correlation
functions would suggest that both liquids share an effective
low-energy description in terms of the same Luttinger liquid, the
different nature of the quasiparticle excitations defies this
intuition. Our interpretation in terms of nonlocal quasiparticle
excitations could also lead to a better understanding of related
models with long-range interactions \cite{Bauer2012,Mattioli2013}.

\begin{acknowledgments}

We acknowledge fruitful discussions with T.~Vekua and A.~Rapp.

\end{acknowledgments}



\end{document}